\documentclass[graybox]{svmult}

\usepackage{type1cm}        
\usepackage{makeidx}         
\usepackage{graphicx}        
\usepackage{multicol}        
\usepackage[bottom]{footmisc}

\usepackage{newtxtext}       %
\usepackage{newtxmath}       
\usepackage{graphicx}
\usepackage{capt-of}
\usepackage{booktabs}
\usepackage{varwidth}

\makeindex             

\begin{document}

\title*{Finite element modelling of in-stent restenosis}
\author{Kiran Manjunatha$^{*}$, Marek Behr, Felix Vogt, Stefanie Reese}

\authorrunning{Manjunatha, K., Behr, M., Vogt, F., Reese, S. }

\institute{$^{*}$ corresponding author\\
\\
Manjunatha, K., Reese, S. \at RWTH Aachen University, Institute of Applied Mechanics, Aachen, Germany\\ \email{kiran.manjunatha@ifam.rwth-aachen.de,  stefanie.reese@ifam.rwth-aachen.de}\\
\and	Behr, M. \at RWTH Aachen University, Chair for Computational Analysis of Technical Systems, Aachen,\\ Germany \\ \email{behr@cats.rwth-aachen.de}
\and	Vogt, F. \at RWTH Aachen University, Department of Cardiology, Pulmonology, Intensive Care and Vascular Medicine, Aachen, Germany\\ \email{fvogt@ukaachen.de}\\
\\
\textit{Submitted for publication in a ``Festschrift'' for Peter Wriggers on December 1,2020.}}
%
%
\maketitle
\vspace{-1.5in}
\hrulefill\\
\abstract{
From the perspective of coronary heart disease, the development of stents has come significantly far in reducing the associated mortality rate, drug-eluting stents being the epitome of innovative and effective solutions. Within this work, the intricate process of in-stent restenosis is modelled considering one of the significant growth factors and its effect on constituents of the arterial wall. A multiphysical modelling approach is adopted in this regard. Experimental investigations from the literature have been used to hypothesize the governing equations and the corresponding parameters. A staggered solution strategy is utilised to capture the transport phenomena as well as the growth and remodeling that follows stent implantation. The model herein developed serves as a tool to predict in-stent restenosis depending on the endothelial injury sustained and the protuberance of stents into the lumen of the arteries.}
\vspace{0.1in}
\vspace{-0.25in}
\noindent\hrulefill
\section{Introduction}
\label{sec:intro}
Coronary heart disease (CHD) is one of the most prominent causes of mortality and affected 126.5 million lives worldwide as of 2017. CHD is characterised by constriction of the coronary artery arising due to the build-up of plaque which consists of lipids, calcifications, and macrophages. This condition, termed atherosclerosis, leads to the narrowing of pathways for blood flow as well as the loss of elasticity of the arterial wall. Percutaneous coronary intervention (PCI) is the process of placing reinforcement structures called stents within these constricted blood vessels to normalize the blood flow. Unfortunately, this interventional procedure is associated with the risk of in-stent restenosis and stent thrombosis. Drug-eluting-stents have emerged in recent years as the most viable option for revascularisation, significantly reducing the risks associated with PCI. But the mechanisms involved in the restenotic process remain incompletely understood. Quantifying the probability and level of restenosis via simulation of the underlying mechanisms will help in addressing the risks more precisely.

Several computational approaches have been developed in this regard. A multi-scale framework involving finite element models and agent-based models has been conceptualised in \cite{zahedmanesh2012}, where the restenotic process was unidirectionally coupled to stresses on the arterial walls. A more recent development on similar grounds has been made in \cite{li2019}, incorporating bidirectional coupling between finite element and agent-based models. A general continuum-based model describing growth and remodelling of soft tissues \cite{humphrey2002}, considering the evolution of constituents in the arterial wall using the constrained-mixture theory, also serves as a computational tool for restenosis prediction. With temporal averaging, a homogenised constrained-mixture model has been developed in \cite{cyron2016} as an extension to the constrained-mixture theory. Alternatively, highly resolved transport phenomena occurring in the arterial wall have also been utilised to model the pathophysiology of vascular diseases. Venous neointimal hyperplasia has been quantified considering the inherent pathogenic mechanisms involving growth factors in \cite{budu2008}. A coupled multi-physical approach for quantifying atherosclerosis has been described in \cite{thon2018}, wherein the mechanics of bloodflow, mass transport and arterial wall mechanics  have been unified under a single framework. An in-stent restenosis predictive model developed on a similar foundation incorporating damage on the arterial wall can be found in \cite{escuer2019}. An analogous attempt is made hereof for the prediction and quantification of the after-effects of stent implantation.

\subsection{Structure of the arterial wall}
The arterial wall is composed of three layers: intima, media, and adventitia. The region where the blood flow occurs is called the lumen. Intima is the innermost layer of the arterial wall, immediately adjacent to the lumen. It is consisted of a monolayer of endothelial cells accompanied by a few layers of smooth muscle cells and extends up to the internal elastic lamina. The media occupies the space between the internal and external laminae and is mainly composed of collagen, elastin and smooth muscle cells. Adventitia, the outermost layer, contains loose connective tissue composed of elastin, collagen and fibroblasts.

\subsection{In-stent restenosis and platelet-derived growth factor}
\label{sec:ISR}
In-stent restenosis refers to the accumulation of new tissue within the intima leading to a diminished cross-section of the lumen post stent implantation. The underlying mechanism is called neointimal hyperplasia. It is a collaborative effect of migration and proliferation of smooth muscle cells (SMC) in the arterial wall, regulated by intricate signalling pathways that are triggered by certain stimuli, either internal or external to the arterial wall. 

The platelet-derived growth factor (PDGF) is a dimer of two peptides linked by a disulfide bond. It has been implicated in vascular remodelling processes, including neointimal hyperplasia, that follow an injury to arterial wall \cite{koyama1994}. This can be attributed to its mitogenic and chemoattractant properties. PDGF is secreted by an array of cellular species namely the endothelial cells, SMC, fibroblasts, macrophages and platelets. 

The stent implantation procedure damages the endothelial monolayer. Also, depending on the arterial overstretch achieved during the implantation, stent struts partially obstruct the blood flow creating vortices in their wake regions. This causes oscillatory wall shear stresses and hence further damages to the endothelium \cite{koskinas2012}. PDGF, which is stored in the alpha-granules of the aggregated platelets at endothelial injury sites, is released into the arterial wall. The presence of PDGF upregulates matrix metalloproteinases (MMP) production in the arterial wall. Extracellular matrix (ECM) is a network of collagen and glycoproteins surrounding the SMC and are degraded in the presence of MMP. SMC, which are usually held stationary by the ECM, are rendered free for migration under the action of MMP. Also, a degraded ECM encourages the proliferation of SMC under the presence of growth factors. The focal adhesion sites created due to cleaved proteins in the ECM assist in the migration of SMC. The direction of migration is influenced by the number of adjacent focal adhesion sites available and the local concentration gradient in PDGF. This directional movement of SMC is termed chemotaxis, and results in the accumulation of the proliferated and migrated SMC in the intima of the arterial wall. A positive feedback loop might occur wherein the migrated SMC create a further obstruction in the blood flow and subsequent upregulation of PDGF. The uncontrolled growth of vascular tissue that follows will eventually lead to a complete blockage of the lumen.

\section{Mathematical modelling}
In-stent restenosis is understood to be a convoluted phenomenon involving multiple constituents. Only the behaviours of key constituents identified from the perspective of the processes described in Sect.\ref{sec:ISR} are modelled. Blood flow in the lumen is not included in the model and the arterial wall is considered to be made of only the intima and media layers.

\subsection{Transport phenomena}
The transport of constituents within the arterial wall is governed by a set of advection-reaction-diffusion equations, the general structure of which for a scalar field $\phi$ is 
\begin{equation}\label{eq:transport}
\displaystyle{\frac{\partial \phi}{\partial t}} = - \underbrace{\nabla \cdot \left(\phi\,\boldsymbol{v}\right)}_{\text{advection}} + \underbrace{\nabla \cdot \left(k\,\nabla \phi\right)}_{\text{diffusion}} + \underbrace{T_s}_{\text{source}} - \underbrace{T_r}_{\text{reaction}},
\end{equation}
where $\boldsymbol{v}$ denotes the velocity of the medium of transport and $k$, the diffusivity of $\phi$ in the medium. To mathematically model in-stent restenosis, PDGF, ECM and SMC are considered to be the key ingredients. Hence the general structure above is adapted to reflect their respective behaviours. The arterial wall is considered to be quasi-static and hence advective terms arising from the movement of the wall are ignored $(\boldsymbol{v} \approx \boldsymbol{0})$.
The symbols and units associated with the transport quantities are declared here for clarity.

\vspace{-0.2in}
\begin{table}[hbt!]
\caption{Transport variables}
\label{tab:1}       
%
%
\begin{tabular}{p{3cm}p{1cm}p{2cm}}
\hline\noalign{\smallskip}
Variable & Symbol & Units  \\
\noalign{\smallskip}\svhline\noalign{\smallskip}
PDGF concentration & $c_{{}_P}$ & mol/mm$^3$\\
ECM density & $\rho_{{}_E}$ & mol/mm$^3$\\
SMC density & $\rho_{{}_S}$ & cells/mm$^3$\\
\noalign{\smallskip}\hline\noalign{\smallskip}
\end{tabular}
\end{table}

\vspace{-0.35in}
\paragraph{Platelet-derived growth factor}
PDGF is assumed to diffuse throughout the arterial wall. Also, it is internalised by the SMCs during their migration and proliferation. The evolution of the concentration of PDGF is thus modelled by the following equation:
\begin{equation}
\displaystyle{\frac{\partial c_{{}_P}}{\partial t}} = \underbrace{\nabla \cdot \left(D_{{}_P}\,\nabla c_{{}_P}\right)}_{\text{diffusion}}  \underbrace{-\,\,\alpha\,\rho_{{}_S}\,c_{{}_P}}_{\text{consumption by SMC}},
 \end{equation}
where $\alpha$ refers to an internalisation coefficient and $\rho_{{}_S}$, the density of SMCs. In comparison to Eq. \ref{eq:transport}, we see that only the terms associated with the diffusive and the reactive phenomena appear here. Although PDGF is produced by other cellular species, the net effect observed is degradation. The diffusion coefficient $D_{{}_P}$ can theoretically vary between positions in the arterial wall but is considered to be constant here.

\paragraph{Extracellular matrix}
The major constituent of ECM is collagen, which is considered to be non-diffusive. SMCs recognize a degraded ECM and synthesize collagen. A source term, in the form of a logistic function, is introduced in this regard and an asymptotic threshold for collagen density $\rho_{{}_{E,th}}$ prescribed. Collagen is degraded by MMP, which in turn is regulated by PDGF. A reactive term is introduced to take care of this degradation. The evolution of ECM density hence reads as follows:
\begin{equation}
\displaystyle{\frac{\partial \rho_{{}_E}}{\partial t}} = \underbrace{\beta\,\rho_{{}_S}\,\left(1 - \displaystyle{\frac{\rho_{{}_{E}}}{\rho_{{}_{E,th}}}}\right)}_{\text{synthesis by SMC}}  \underbrace{-\,\,\gamma\,c_{{}_P}\,\rho_{{}_{E}}}_{\text{degradation due to MMP}}.
\end{equation}
Here, $\beta$ and $\gamma$ refer to synthesis and degradation rate coefficients respectively.
 
\paragraph{Smooth muscle cells}
The migration of SMC under the influence of PDGF is modelled using a chemotaxis term \cite{keller1971}, which can be interpreted as a pseudo-advective term wherein the velocity of the medium is replaced by the gradient in the ECM density. This corresponds to the fact that higher degradation in the ECM results in higher focal adhesion sites for SMC migration. Additionally, a source term is introduced to model the proliferation of SMC under the presence of PDGF. Hence the evolution of the SMC density is prescribed using the equation:
\begin{equation}
\displaystyle{\frac{\partial \rho_{{}_S}}{\partial t}} = \underbrace{\nabla \cdot \left(\chi\,c_{{}_P}\left(1 - \displaystyle{\frac{\rho_{{}_{E}}}{\rho_{{}_{E,th}}}} \right)\,\rho_{{}_S}\,\nabla \rho_{{}_E}\right)}_{\text{chemotaxis due to ECM degradation}} + \underbrace{\kappa\,c_{{}_P}\,\left(1 - \displaystyle{\frac{\rho_{{}_{E}}}{\rho_{{}_{E,th}}}} \right)\,\rho_{{}_S}}_{\text{proliferation due to PDGF}}.
\end{equation}
The chemotactic sensitivity $\chi$ and the proliferation constant $\kappa$ are scaled by the concentration of PDGF as well as the logistic coefficient dependent on the ECM density. This takes care of the fact that the migration and proliferation effects increase with an increase in ECM degradation, in the presence of PDGF.

\subsection{Arterial wall mechanics}
The continuum mechanical description of in-stent restenosis requires first the definition of the kinematics of growth and remodelling of the arterial wall, incorporating relevant quantities from the transport phenomena previously described. Constitutive equations hence based on the kinematics, in combination with the balance of linear momentum, will complete the mathematical model description.

\subsection{Kinematics}
A deformation map $\boldsymbol{\varphi}$ of a particle at position $\boldsymbol{X}$ in the reference configuration $\Omega_0$ at time $t_0$ to its position $\boldsymbol{x}$ in the current configuration  $\Omega$ at time $t$ is provided by the deformation gradient $\boldsymbol{F} = \nabla_{{}_{\boldsymbol{X}}} \boldsymbol{\varphi}(\boldsymbol{X},t)$. The right Cauchy-Green tensor is further defined as $\boldsymbol{C} = \boldsymbol{F}^T \,\boldsymbol{F} $.

For a description of growth, a multiplicative decomposition of the deformation gradient is adopted. Assumption of an intermediate incompatible configuration which achieves a locally stress-free state forms the basis of this split \cite{himpel2005,rodriguez1994}. An additional elastic deformation is needed to ensure the compatibility of the total deformation. Hence the total deformation gradient will be $\boldsymbol{F} = \boldsymbol{F}_e\,\boldsymbol{F}_g$ and the elastic right Cauchy-Green tensor reads $\boldsymbol{C}_e = \boldsymbol{F}_e^T\,\boldsymbol{F}_e$. 

The growth deformation gradient is specified as  $\boldsymbol{F}_g = \vartheta\,\boldsymbol{I}$, under the assumption of isotropic growth.
Hence the compatible elastic deformation gradient will be $\boldsymbol{F}_e = \vartheta^{-1}\,\boldsymbol{F}.$

To calculate the growth stretch $\vartheta$, it is hypothesised that an increase in volume at a point due to growth at any time $t$ is proportional to the additional SMC transported to that point. A similar idea is postulated for aggregation of foam cells in \cite{thon2018}. If $v_s$ is the volume occupied by a single SMC and $N_s$ the number of additional SMC at a point, then
$\Delta V_g = v_s\,N_s$. $v_s$ can be prescribed via the definition of the SMC density of a healthy artery $\rho_{{}_{S,h}}$, using $v_s = (\rho_{{}_{S,h}})^{-1}$. If $V_g$ is the volume in the intermediate growth configuration $\Omega_g$ at time $t$ and $V_0$ is that in the reference  configuration $\Omega_0$, then we can arrive at an expression for $\vartheta$ through the following exercise.
\begin{eqnarray}
\Delta V_g = V_g - V_0 &=& (\rho_{{}_{S,h}})^{-1}\,N_s\\
\displaystyle{\int_{\Omega_g}} 1\,dv_g -  \displaystyle{\int_{\Omega_0}} 1\,dV &=& (\rho_{{}_{S,h}})^{-1}\,   \displaystyle{\int_{\Omega}} (\rho_{{}_{S}} - \rho_{{}_{S,h}})\,dv\\
\nonumber
\end{eqnarray}
Pulling back the quantities to the initial configuration,
\begin{equation}
\displaystyle{\int_{\Omega_0}} J_g \,dV -  \displaystyle{\int_{\Omega_0}} 1\,dV = (\rho_{{}_{S,h}})^{-1}\,   \displaystyle{\int_{\Omega_0}} J\,(\rho_{{}_{S}} - \rho_{{}_{S,h}})\,dV \label{volume balance}
\end{equation}  
Since Eq. \ref{volume balance} has to hold true locally,
\begin{eqnarray}
J_g  -  1 &=& (\rho_{{}_{S,h}})^{-1}\,J\,(\rho_{{}_{S}} - \rho_{{}_{S,h}})\\
\implies J_g  &=& 1 + \displaystyle{J\,\left(\frac{\rho_{{}_{S}}}{\rho_{{}_{S,h}}} - 1\right)}.\\
\nonumber
\end{eqnarray}
If $d$ is the dimensionality of the problem, then $J_g = \det (\boldsymbol{F}_g)=\vartheta^d$. Therefore,
\begin{equation}\label{growth factor}
\vartheta = \left(1 + \displaystyle{J\,\left(\frac{\rho_{{}_{S}}}{\rho_{{}_{S,h}}} - 1\right)}\right)^{1/d}.
\end{equation}

\subsection{Hyperelastic constitutive model}
The constitutive model is derived from the hyperelastic models presented in \cite{gasser2006,holzapfel2000,nolan2014}, wherein the arterial wall is assumed to be composed of two helices of collagen fibres, with respective helix angles $\theta^a_i$, embedded in an isotropic ground substance. The associated Helmholtz free energy per unit volume is hence split into an isotropic and an anisotropic part, both of which are dependent on the elastic right Cauchy-Green tensor. 
\begin{eqnarray}
\psi(\boldsymbol{C}_e, \boldsymbol{H}_1, \boldsymbol{H}_2) &=& \psi_{iso}(\boldsymbol{C}_e) + \psi_{ani}(\boldsymbol{C}_e, \boldsymbol{H}_1, \boldsymbol{H}_2)  \\
\psi_{iso}(\boldsymbol{C}_e) &=& \displaystyle{\frac{\mu^a}{2}}\left(\text{tr}\,\boldsymbol{C}_e - 3\right) - \mu^a\,\text{ln}\,J_e + \displaystyle{\frac{\Lambda^a}{4}}\left(J_e^2 - 1 - 2\,\text{ln}\,J_e \right)\\
\psi_{ani}(\boldsymbol{C}_e, \boldsymbol{H}_1, \boldsymbol{H}_2) &=& \displaystyle{\frac{k_1}{2k_2}}\sum_{i=1,2} \left(\text{exp}\left[k_2\langle E_i\rangle ^2 \right]-1\right)\\
\nonumber
\label{eqarray constitutive model}
\end{eqnarray}
This stems from the hypothesis that in the stress-free incompatible growth configuration, even the collagen fibres grow due to the synthesis by SMC and hence no residual stresses arise in the incompatible state. The dependence of $\psi$ on $\boldsymbol{C}_e$ embeds its dependence on $\boldsymbol{F}_g$ via 
\begin{equation}\label{rcg_e}
\boldsymbol{C}_e = \boldsymbol{F}_g^{-T}\, \boldsymbol{C}\, \boldsymbol{F}_g^{-1} = \vartheta^{-2} \boldsymbol{C}.
\end{equation}

The generalised structure tensors $\boldsymbol{H}_i$ are constructed based on the local collagen orientations in the reference configuration $\boldsymbol{a}_{0i}$ using 
\begin{equation}
\boldsymbol{H}_i = \kappa^a\,\boldsymbol{I} + \left(1 - 3\,\kappa^a \right)\,\boldsymbol{a}_{0i} \otimes \boldsymbol{a}_{0i}\,\,,
\end{equation}
where $\kappa^a$ is a dispersion parameter. The Green-Lagrange strain like parameter $E_i$ is calculated utilising the relationship $E_i = \boldsymbol{H}_i : \boldsymbol{C}_e - 1,$ where the definition of scalar product of second order tensors $\boldsymbol{A}:\boldsymbol{B} = A_{ij}\,B_{ij}$ (Einstein summation convention) is applied.

The first Piola-Kirchhoff stress tensor is deduced from the Helmholtz free energy function using
\begin{equation}
\boldsymbol{P} = \displaystyle{\frac{\partial \psi}{\partial \boldsymbol{F}}} = \displaystyle{\frac{\partial \psi}{\partial \boldsymbol{C}_e}}:\displaystyle{\frac{\partial \boldsymbol{C}_e}{\partial \boldsymbol{F}}}.
\end{equation}  
Using Eq. \ref{rcg_e}, it can be deduced that 
\begin{equation}
\displaystyle{\frac{\partial \boldsymbol{C}_e}{\partial \boldsymbol{F}}} = \displaystyle{\frac{1}{\vartheta^2}}\,\displaystyle{\frac{\partial \boldsymbol{C}}{\partial \boldsymbol{F}}}.
\end{equation}
Finally, the balance of linear momentum governs the quasi-static equilibrium of the arterial wall structure.
\begin{equation}
\nabla \cdot \boldsymbol{P} + \boldsymbol{b} = \boldsymbol{0} 
\end{equation}
\section{Numerical methods}
The Galerkin weak forms of the governing differential equations are arrived at.
\begin{eqnarray}
\displaystyle{\int_{\Omega}\frac{\partial c_{{}_{P}}}{\partial t}}\,\delta c_{{}_{P}}\,dv &=& \displaystyle{\int_{\Gamma} (J_{{}_{P}} \cdot \boldsymbol{n})\,\delta c_{{}_{P}}\,da} - \displaystyle{\int_{\Omega} (\nabla \delta c_{{}_{P}} \cdot D_{{}_P}\nabla c_{{}_{P}}})\,dv \nonumber\\
 &-& \displaystyle{\int_{\Omega} \alpha\,\rho_{{}_{S}}\,c_{{}_{P}} }\,\delta c_{{}_P}\,dv\\
\displaystyle{\int_{\Omega}\frac{\partial \rho_{{}_{E}}}{\partial t}}\, \delta \rho_{{}_{E}}\,dv &=&  \displaystyle{\int_{\Omega}\,\beta\, \rho_{{}_{S}} \left(1 - \displaystyle{\frac{\rho_{{}_{E}}}{\rho_{{}_{E,th}}}}\right)\, \delta\rho_{{}_E}} \,dv - \displaystyle{\int_{\Omega}\,\gamma c_{{}_{P}} \rho_{{}_{E}}\, \delta\rho_{{}_{E}}} \,dv\\
\displaystyle{\int_{\Omega} \frac{\partial \rho_{{}_{S}}}{\partial t}}\,\delta \rho_{{}_S} \,dv &=& \displaystyle{\int_{\Gamma} (J_{{}_{S}} \cdot \boldsymbol{n})\,\delta \rho_{{}_{S}}\,da} - \displaystyle{\int_{\Omega}\left(\nabla\,\delta\rho_{{}_{S}} \cdot \chi\,c_{{}_P}\,\left(1 - \displaystyle{\frac{\rho_{{}_{E}}}{\rho_{{}_{E,th}}}}\right) \,\rho_{{}_{S}} \,\nabla \rho_{{}_{E}}\right)} \,dv \nonumber\\
&+& \displaystyle{\int_{\Omega} \kappa\,c_{{}_{P}}\, \rho_{{}_{S}} \left(1 - \displaystyle{\frac{\rho_{{}_{E}}}{\rho_{{}_{E,th}}}}\right)}\,\delta\rho_{{}_S}\,dv\\
\displaystyle{\int_{\Omega_0}}\, \boldsymbol{P}:\delta\boldsymbol{F}\, dV &=& \displaystyle{\int_{\Gamma_0}}\,(\boldsymbol{T}\cdot\delta\boldsymbol{u})\,dA\label{wf:blm}\\
\nonumber
\end{eqnarray}
Here, $J_{{}_P}$ and $J_{{}_S}$ denote the PDGF and SMC fluxes across the boundary, $\boldsymbol{n}$ the normal on the boundary and $\boldsymbol{T}$ the traction on the boundary. Body forces are considered absent. 

The transport equations are temporally discretised using the semi-implicit Backward-Euler method. Semi-implicitness is attributed to the fact that in each of the weak forms, variables other than the ones discretised in time are carried over from the previous time step. Spatial discretisation is performed using linear finite elements. A decoupled set of linear equations for the time-step updation of transport variables are hence obtained.
\begin{eqnarray}
\left[\boldsymbol{M} + \Delta t \,\boldsymbol{L} + \Delta t \,\boldsymbol{P} \right] \boldsymbol{c}_{{}_{P}}^{n+1} &=& \boldsymbol{M}\boldsymbol{c}_{{}_{P}}^{n}
\\
\left[\boldsymbol{M} + \Delta t \,\boldsymbol{T} \right] \boldsymbol{\rho}_{{}_{E}}^{n+1} &=& \boldsymbol{M}\boldsymbol{\rho}_{{}_{E}}^{n} + \boldsymbol{R}
\\
\left[\boldsymbol{M} + \Delta t \,\boldsymbol{K}  - \Delta t \,\boldsymbol{Q}\right] \boldsymbol{\rho}_{{}_{S}}^{n+1} &=& \boldsymbol{M}\boldsymbol{\rho}_{{}_{S}}^{n}\\
\nonumber
\end{eqnarray}
\vspace{-0.3in}\\
Definitions of matrices in the equations above can be found in the appendix. The diffusion and advection operators within the finite element setting, when discretised with linear finite elements, display instabilities across steep gradients in the transport quantities. This leads to unphysical negative concentrations and densities at the diffusion and advection fronts. To treat this, artificial discrete diffusive fluxes are introduced, which result in over smoothened solution profiles. Hence antidiffusive fluxes, limited using the Zalesak limiter, are utilised to preserve local extrema \cite{kuzmin2002,strehl2013}. 

The weak form of the balance of linear momentum in Eq. \ref{wf:blm} is spatially discretised using the same linear finite elements as defined for the transport problems, and is solved iteratively using the Newton-Raphson method. The residual and its linearization, prior to discretisation, required for the iterative updates are determined as follows.
 
\begin{eqnarray}
g_{{}_u} &=& \displaystyle{\int_{\Omega_0}}\, \boldsymbol{P}:\delta\boldsymbol{F}\, dV - \displaystyle{\int_{\Gamma_0}}\,(\boldsymbol{T}\cdot\delta\boldsymbol{u})\,dA\\
\Delta g_{{}_u} &=& \displaystyle{\int_{\Omega_0}} \delta\boldsymbol{F} : (\mathcal{A}:\Delta \boldsymbol{F})\,dV\\
\nonumber
\end{eqnarray}
The fourth order tensor $\mathcal{A}$ is computed using
\begin{equation}
\mathcal{A} = \displaystyle{\frac{\partial \boldsymbol{P}}{\partial \boldsymbol{F}} + \frac{\partial \boldsymbol{P}}{\partial \vartheta} \otimes \frac{\partial \vartheta}{\partial \boldsymbol{F}}}.
\end{equation}

\paragraph{Coupling}
The transport problems are solved in the current configuration $\Omega$ while the balance of linear momentum is solved in the reference configuration $\Omega_0$. A staggered solution strategy is carried out to achieve coupling. The transport problem provides the SMC density to the structural problem while the structural problem updates the current configuration $\Omega$ based on the SMC density. An incremental form of Eq. \ref{growth factor} is applied in this regard as follows:
\begin{equation}
\vartheta = \vartheta^- \left(1 + \displaystyle{\frac{J}{J^-}\,\left(\frac{\rho_{{}_{S}}}{\rho_{{}_{S,h}}} - 1\right)}\right)^{1/d},
\end{equation}
wherein the quantities $(\,\,)^-$ refer to those from the previous time step. At the end of solution of the structural problem, the transport quantities $\phi$ are updated at the Gauss-points using \(\phi^* = (\,J^{-}/J)\,\phi\) and extrapolated to the nodes. The staggered solution strategy provides flexibility in case different spatial discretisations are absolutely necessary for the systems being coupled.

\section{Numerical investigation}
A longitudinal axisymmetric section of an artery is considered fixed on either ends. A Gaussian distribution of PDGF concentration consisting of three distinct peaks is prescribed in the domain to mimic a PCI injury in the arterial wall. ECM and SMC densities ($\rho_{{}_{E,0}}$ and $\rho_{{}_{S,0}}$) are uniformly prescribed to be those of a healthy artery. Zero flux boundary conditions are prescribed for all the transport quantities on all boundaries. Some of the parameters are deduced from literature and the others tuned to predict the physics qualitatively. 

\begin{table}[htb!]

\begin{varwidth}[b]{0.3\linewidth} 
\centering     
\begin{tabular}{p{2cm}p{4.0cm}}
\hline\noalign{\smallskip}
Parameter & Value [units]\\
\noalign{\smallskip}\svhline\noalign{\smallskip}
$D_{{}_P}$ & $0.01$ [mm$^2$/day]\\
$\alpha$ & $1.0 \times 10^{-13}$ [mm$^3$/cell/day]  \\
$\beta$ & $5.0 \times 10^{-8}$ [mol/cell/day]\\
$\gamma$ & $5.0 \times 10^{17}$ [mm$^3$/mol/day]\\
$\chi$ & $1.0 \times 10^{19}$ [mm$^5$/cell/day]\\
$\kappa$ & $1.0 \times 10^{-2}$ [mm$^3$/mol/cell/day]\\
$\rho_{{}_{E,0}}$ & $7.0 \times 10^{-9}$ [mol/mm$^3$]\\
$\rho_{{}_{E,th}}$ & $1.1 \times \rho_{{}_{E,0}}$\\
$\rho_{{}_{S,0}}$ & $3.16 \times 10^{6}$ [cells/mm$^3$]\\
$\mu^a$ & $0.02$ [M\,Pa]\\
$\lambda^a$ & $10$ [M\,Pa]\\
$k_1$ & $0.112$ [M\,Pa]\\
$k_2$ & $20.61$ [-]\\
$\kappa^a$ & $0.1$ [-]\\
$\theta^a_i$ & $41$ [$^\circ$]\\
\noalign{\smallskip}\hline\noalign{\smallskip}
\end{tabular}
\caption{Parameter set}
\label{tab:parameter set} 
\end{varwidth}%
\hfill
\begin{minipage}[b]{0.5\linewidth}
\vspace{-1in}
\centering
\includegraphics[scale=.4]{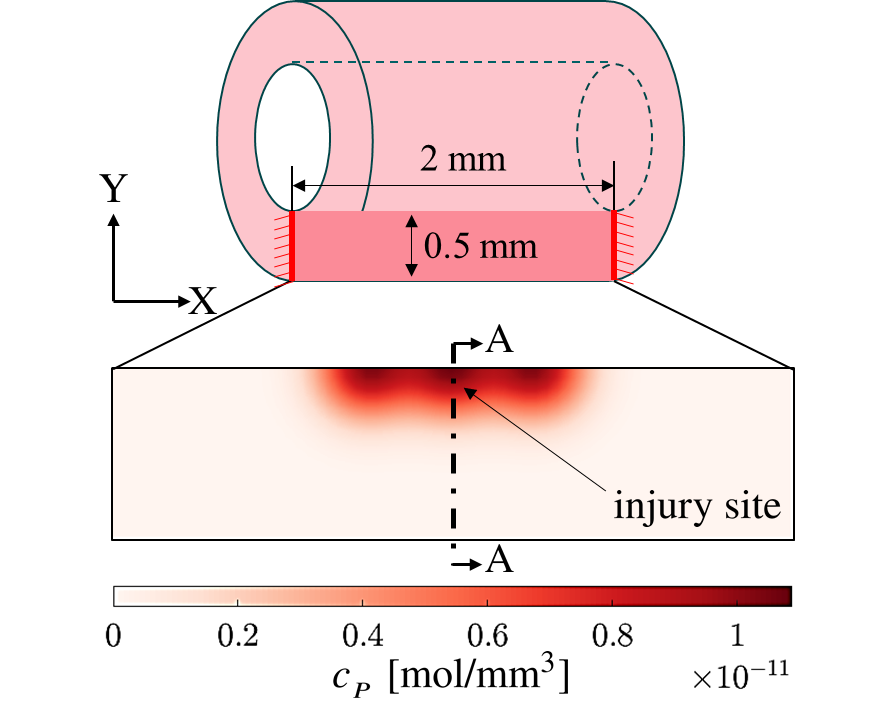}
\captionof{figure}{Problem setup}
\end{minipage}
\end{table}

Transport simulation is run for a time period of 1 day to perform the mesh and time step convergence analyses since this dictates the spatial and temporal discretisations. The SMC density plots along section A-A are used for the analyses.

\begin{figure}
\centering
\begin{minipage}{0.55\textwidth}
\includegraphics[scale=.35]{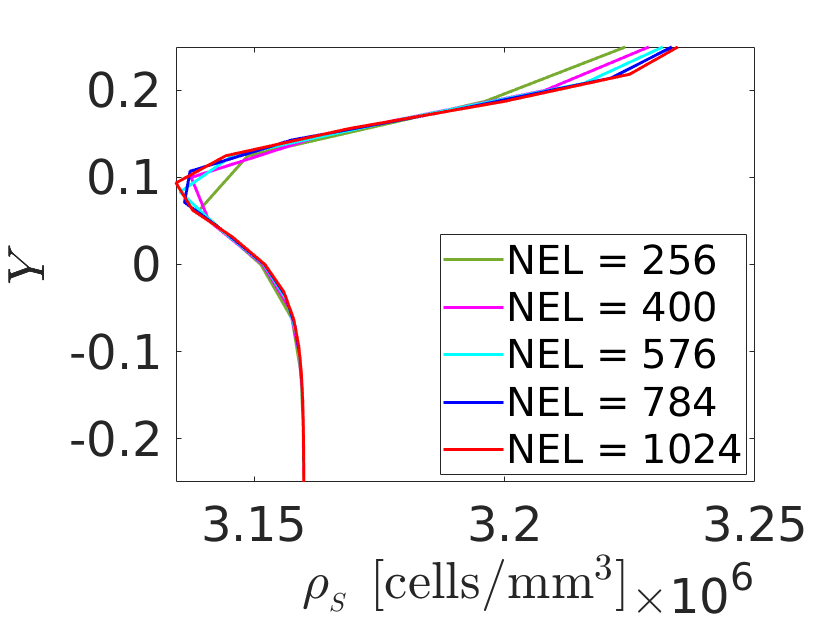}
\caption{mesh convergence (A-A)}
\label{fig:problem setup} 
\end{minipage}
\begin{minipage}{0.4\textwidth}
\includegraphics[scale=.35]{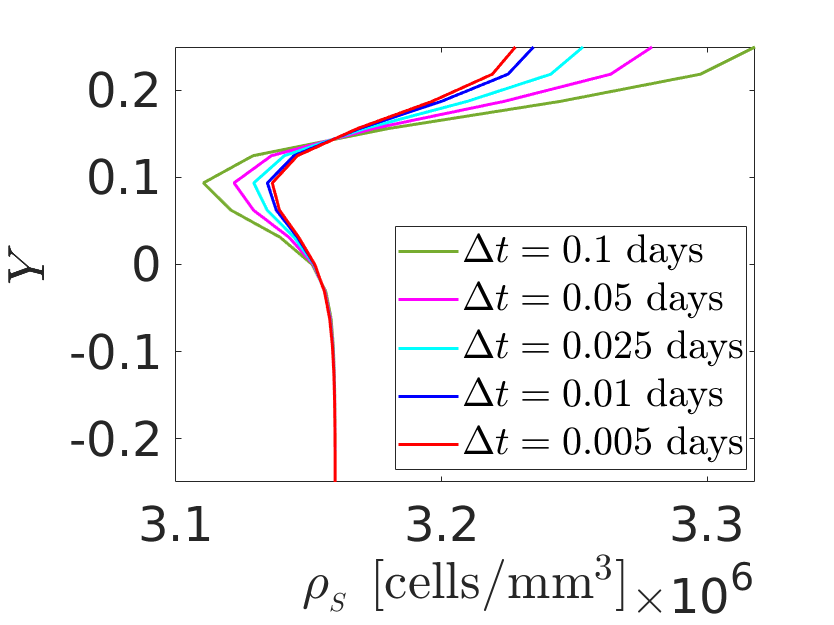}
\caption{time-step convergence (A-A)}
\label{fig:problem setup}
\end{minipage}
\end{figure}

\begin{figure}[htb!]
\centering
\label{gp_evo}
\includegraphics[width=\textwidth]{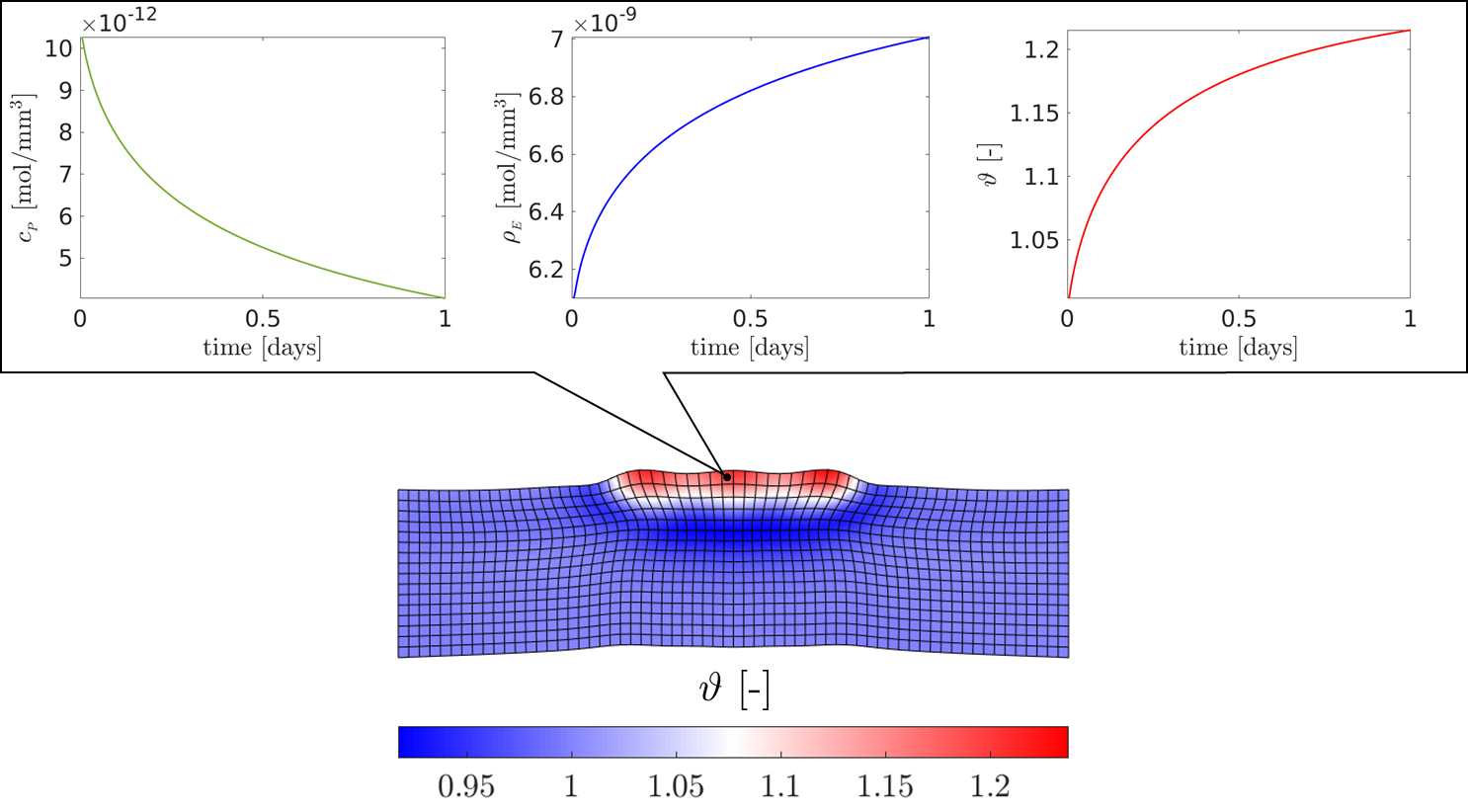}
\caption{deformed arterial wall and evolution of quantities of interest}
\end{figure}

Using the converged discretisations, the coupled simulation is run for a period of 1 day and the results examined.
It is expected that the growth ceases when the ECM is healed and PDGF is completely internalized. The evolutions of quantities at a Gauss-point of interest reflect this as shown in Fig. \ref{gp_evo}. Contraction is observed in some regions of the arterial wall which can be attributed to the migration of SMC away from these regions.

\section{Conclusion and outlook}
The developed model is able to successfully capture the growth induced in the arterial wall due to mechanisms brought about by PDGF. It is to be noted that the model does not take into account the loss and healing processes of the endothelium. In reality, the restenotic process does not cease until the endothelium is completely healed and there is no further influx of PDGF. Wall shear stress dependent influx of PDGF can be prescribed as a boundary condition to achieve the aforementioned effect. The transforming growth factor (TGF)-$\beta$ is another significant constituent responsible for modulation of growth in vasculatures. It considerably affects the proliferative response of SMCs and hence needs to be considered in the framework developed in this work. Furthermore, it remains to be seen how sensitive the model is to the set of model parameters and which of those can be determined experimentally.

Future work shall therefore entail consideration of additional wall constituents that influence the restenotic process, physiological experimentation, parameter sensistivity evaluation, and finally arriving at a high-fidelity in-silico model which aides in examining the efficacy of drugs intended to be used on drug-eluting stents.

\begin{acknowledgement}
The financial support of DFG for the projects 'Drug-eluting coronary stents in stenosed arteries: medical investigation and computational modelling' (RE 1057 / 44-1) and 'Modelling of Structure and Fluid-Structure Interaction during Tissue Maturation in Biohybrid Heart Valves' (RE 1057 / 45-1) is gratefully acknowledged.
\end{acknowledgement}
\vspace{-0.4in}
\section*{Appendix}
\addcontentsline{toc}{section}{Appendix}
Element matrices in the linearised decoupled system of transport equations are listed here. $\boldsymbol{N}$ refers to the shape function vector associated with linear finite elements.
\begin{align*}
\begin{array}{r@{\mskip\thickmuskip}l}
\boldsymbol{M}^e &= \displaystyle{\int_{\Omega_e}}\, \boldsymbol{N}\,\boldsymbol{N}^T\,dv
\nonumber
\\
\boldsymbol{L}^e &= \displaystyle{\int_{\Omega_e}}\,D_{{}_{P}} \nabla \boldsymbol{N}\,\nabla \boldsymbol{N}^T\,dv
\nonumber
\\
\boldsymbol{P}^e &= \displaystyle{\int_{\Omega_e}}\,\alpha \rho_{{}_{S}}^n \boldsymbol{N}\,\boldsymbol{N}^T\,dv
\nonumber
\end{array}
\quad
\begin{array}{r@{\mskip\thickmuskip}l}
\boldsymbol{T}^e &= \displaystyle{\int_{\Omega_e}}\left(\beta\,\displaystyle{\frac{\rho_{{}_{S}}^n}{\rho_{{}_{E,th}}}} + \gamma\,c_{{}_P}^n\right)\boldsymbol{N}\,\boldsymbol{N}^T\,dv
\nonumber
\\
\boldsymbol{K}^e &=  \displaystyle{\int_{\Omega_e}}\left[ \chi\,c_{{}_{P}}^n\left(1 - \displaystyle{\frac{\rho_{{}_{E}}^n}{\rho_{{}_{E,th}}}}\right)\nabla \boldsymbol{N}\, \nabla \rho_{{}_1}^n \boldsymbol{N}^T\right]\,dv \nonumber
\\
\boldsymbol{Q}^e &= \displaystyle{\int_{\Omega_e}}\,\kappa\,c_{{}_{P}}^n \left(1 - \displaystyle{\frac{\rho_{{}_{E}}^n}{\rho_{{}_{E,th}}}}\right)\boldsymbol{N}\,\boldsymbol{N}^T\,dv
\nonumber
\\
\boldsymbol{R}^e &= \displaystyle{\int_{\Omega_e}} \beta\,\rho_{{}_S}^n\,\boldsymbol{N}\,dv
\nonumber
\end{array}
\end{align*}

\vspace{-0.3in}
%
%

\end{document}